# Empirical results for pedestrian dynamics and their implications for cellular automata models

Andreas Schadschneider [a] and Armin Seyfried [b]

[a] Institut für Theoretische Physik, Universität zu Köln, 50937 Köln, Germany
[b] Jülich Supercomputing Centre, Forschungszentrum Jülich GmbH, 52425 Jülich, Germany

`as@thp.uni-koeln.de, a.seyfried@fz-juelich.de`

**Abstract**

A large number of models for pedestrian dynamics have been developed over the years. However, so far not much attention has been paid to their quantitative validation. Usually the focus is on the reproduction of empirically observed collective phenomena, as lane formation in counterflow. This can give an indication for the realism of the model, but practical applications, e.g. in safety analysis, require *quantitative* predictions. In this chapter we discuss the current experimental situation, especially for the fundamental diagram which is the most important quantity needed for calibration. In addition we consider the implications for the modelling based on cellular automata. As specific example the floor field model is introduced. Apart from the properties of its fundamental diagram we discuss the implications of an egress experiment for the relevance of conflicts and friction effects.

# 1. Introduction

In recent years a large number of models for the simulation of pedestrian dynamics has been proposed, some of them being quite successful in providing a realistic description of a variety of different situation. In contrast, the empirical situation is much less satisfactory. Not much experimental data are available and if they are, they are often unreliable. This is reflected in the fact that the data are sometimes even contradictory, see e.g. (Schadschneider et al., 2009), even for the simplest scenarios. This might be one of the reasons why so far not many models have been tested quantitatively by comparing with empirical data. Instead the reproduction of collective phenomena like lane formation, oscillations at bottlenecks or pattern formation at intersections has been used as a criterion to judge the realism of the models.

Therefore there have been only a few attempts to calibrate and validate models of pedestrian dynamics properly. The application of models in the area of safety planning is somewhat limited or has to be taken with a grain of salt. A first important step to improve the current state of affairs would be to obtain reliable empirical data. This is an essential first step and would form the basis for validation and calibration. Only then one can make even quantitative predictions based on computer simulations.

Perhaps the most important characteristic of pedestrian dynamics is the fundamental diagram, i.e. the relation between pedestrian flow and its density. It is of obvious importance for the dimensioning of pedestrian facilities. Furthermore it is associated with many self-organization phenomena, like the formation of lanes or the occurrence of jams. However, even for this basic quantity the current situation is largely confusing (see Sec. 2).

In most models, pedestrians are considered to be autonomous mobile agents, hopping particles in a cellular automaton or self-driven particles in continuous space. These model classes form the basis for sophisticated multi-agent simulations. It is worth mentioning that in physics usually "multi-agent model" is taken as a synonym for "microscopic model". Usually one takes into account that a model should be a) as realistic as possible and b) flexible enough for different realistic applications. Point b) is generically realized by multi-agent approaches which provide an environment to include the infrastructure, visualization etc. In this spirit we will focus here on point a), the realism of the modelling approach. This is intimately related to the qualitative and quantitative comparison with empirical data.

In Sec. 2 we compare existing various experimental data and specifications from the literature and discuss the observed discrepancies. The focus is on the fundamental diagram and the flow through a bottleneck.

In Sec. 3 we will review the basic modelling approaches focusing on cellular automata models. We present the floor field model, discuss the characteristics of this approach and discuss quantitative results obtained from computer simulations, especially for the fundamental diagram. By introducing the concept of "friction" the model is able to reproduce results from a large-scale evacuation experiment.

## 2. Empirical results and validation

### 2.1. Principles of validation

Before any model is used in applications, especially in sensitive areas like safety analysis, it should be properly validated and calibrated (if reliable quantitative results are needed). But which principles should be used in the validation procedure? So far it appears that there is no consensus on this point and that everybody comes up with his/her own criteria. Often these appear to be somewhat biased by the performance of the own favourite models and one tends to prefer methods where the own model fairs better.

Regarding validation, one could distinguish between "qualitative" versus "quantitative" and "macroscopic" versus "microscopic" validation procedures. Qualitative means that certain collective phenomena like lane formation or the formation of jams are reproduced qualitatively. Quantitative validation in contrast would test whether in case of lane formation the quantitative relation between velocity and density or in case of jam formation the value of the jam density is reproduced correctly. Regarding quantitative validation, one could distinguish between "macroscopic" and "microscopic" observables used for the procedure. Macroscopic means that the observable considered is a mean value over time or space. Microscopic validation in contrast would test more individual properties like individual velocities and their distribution at a certain density or properties of single trajectories, like the curvature.

For quantitative macroscopic validation it is important to note that system sizes as well as measuring methods have to be the same for comparison of experimental data with simulation results. Experimental data of pedestrian flow are often connected with inhomogeneities in space and time, finite size effects and non-equilibrium conditions.

Ideally the validation procedure should guarantee that the model works in very general settings, not just in the scenarios tested. How to achieve this is not obvious. For pedestrian dynamics one should try to formulate a number of tests a model should pass. We suggest, as part of these tests, to consider macroscopic trajectories, like the formation of lanes in counterflow and in narrow bottlenecks. Furthermore, qualitative aspects of the fundamental diagrams for strictly one-dimensional motion and at bottlenecks should be reproduced.

The fundamental diagram is the most important characteristic of pedestrian dynamics. Besides its importance for the dimensioning of pedestrian facilities it is associated with every qualitative self-organization phenomenon, like the formation of lanes or the occurrence of jams. However, specifications of various experimental studies, guidelines and handbooks display substantial differences in maximal flow values and the corresponding density as well as the density where the flow vanishes due to overcrowding. Different explanations for these discrepancies have been proposed, ranging from differences between uni- and multidirectional flow and cultural or population effects to psychological factors given by the incentive for the movement. Also the behaviour at bottlenecks is far from being understood, e.g. why the flow can be significantly larger than the maximum of the fundamental diagram.

A validation of models with fundamental diagrams for (quasi-) one-dimensional motion only is certainly not sufficient. Pedestrian dynamics is complex due to its two-dimensional nature. However, it is believed that the behaviour in one-dimensional scenarios can reflect the most relevant aspects of the significant interactions. Nevertheless this should be verified later e.g. by measuring fundamental diagrams for genuine two-dimensional motions.

This program makes only sense if sufficient reliable empirical data are available. Unfortunately this is not the case and the empirical understanding of pedestrian dynamics is far from satisfactory.

## 2.2. Fundamental diagram

The most basic quantities to characterize the collective properties pedestrian (or, more generally, 'particle') motion are the density $\rho$ and flow $J$ (or specific flow per unit width $J_s = J/b$). The relation between these quantities is usually called *fundamental diagram* which already indicates its importance. Due to the hydrodynamic relation $J = \rho v b$, where $v$ is the average velocity, three equivalent forms are used: $J_s(\rho)$, $v(\rho)$ and $v(J_s)$.

In applications the fundamental diagram is a basic input for most engineering methods developed for the design and dimensioning of pedestrian facilities (Predtechenskii and Milinskii, 1978; Fruin, 1971; Nelson and Mowrer, 2002). In the following we will consider only planar facilities like sidewalks, corridors or halls. Other facilities like floors, stairs or ramps are less well studied and the shape of the diagrams can differ from the planar case.

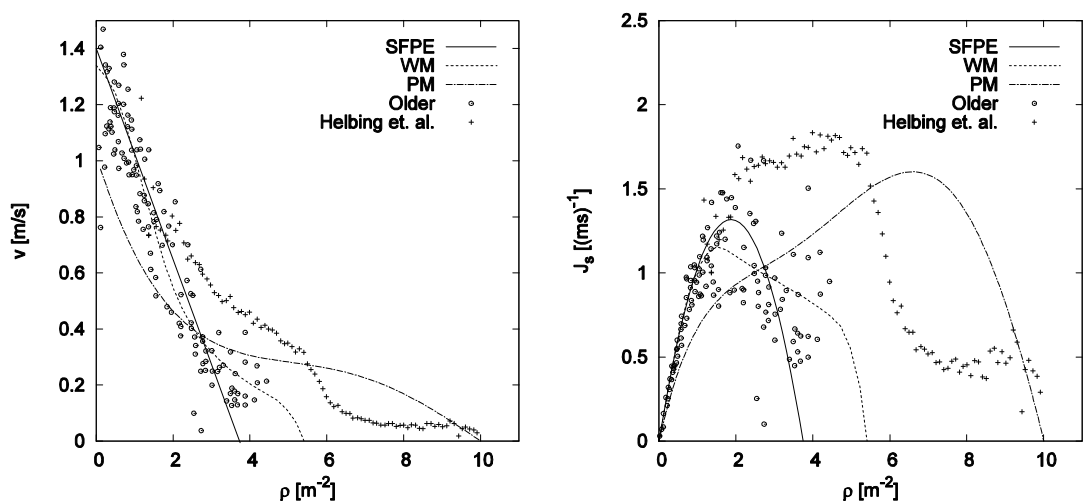

Fig. 1. Fundamental diagrams for pedestrian movement in planar facilities. Lines refer to specifications in planning guidelines PM: (Predtechenskii and Milinskii, 1978); SFPE: (Nelson and Mowrer, 2002) and WM: (Weidmann, 1993). Data points are obtained from experimental measurements (Older, 1968) and (Helbing et al., 2007).

In Fig. 1 fundamental diagrams which are frequently used in planning guidelines are shown. For comparison, results from two selected empirical studies are also included to demonstrate the variance of the data. Natural quantities that can be used to characterize empirical

fundamental diagrams are
- the maximum of the function or capacity $J_{s,max}$ ;
- the density $\rho_c$ where the maximum flow is reached ;
- the density $\rho_0$ where the velocity approaches zero due to overcrowding.

As seen in Fig. 1 the specifications and measurements even for these most basic characteristics disagree considerably:
- $1.2\ (ms)^{-1} < J_{s,max} < 1.8\ (ms)^{-1}$,
- $1.75\ m^{-2} < \rho_c < 7\ m^{-2}$,
- $3.8\ m^{-1} < \rho_0 < 10\ m^{-1}$.

Several explanations for these deviations have been suggested, e.g.
- cultural and population differences (Helbing et al., 2007),
- differences between uni- and multidirectional flow (Navin and Wheeler, 1969; Pushkarev and Zupan, 1975),
- short-ranged fluctuations (Pushkarev and Zupan, 1975),
- influence of psychological factors given by the incentive of the movement (Predtechenskii and Milinskii, 1978) or the type of traffic (commuters, shoppers) (Oeding, 1963).

However, currently no consensus about the relevance of these factors has been reached. For example it is not even clear whether there is a difference between fundamental diagrams obtained from uni-directional and multi-directional flows. Weidmann (Weidmann, 1993) neglects these differences and Fruin (Fruin, 1971) argues that the flows in these situations differ only slightly. However, this disagrees with results of Navin and Wheeler (Navin and Wheeler, 1969) who found a reduction of the flow in dependence of directional imbalances.

This brief discussion clearly shows that up to now there is no consensus even on the basic characteristics of the fundamental diagram or its precise form. Even the origin of the observed discrepancies is still discussed controversially.

Another aspect which plays a role when comparing data from different sources is the fact that in the majority of cases error margins or even fluctuations are not shown. Furthermore, as is is well-known from vehicular traffic, different measurement methods can lead to deviations for the resulting relations (Leutzbach, 1988; Kerner, 2004). This is exemplified in Fig. 2.

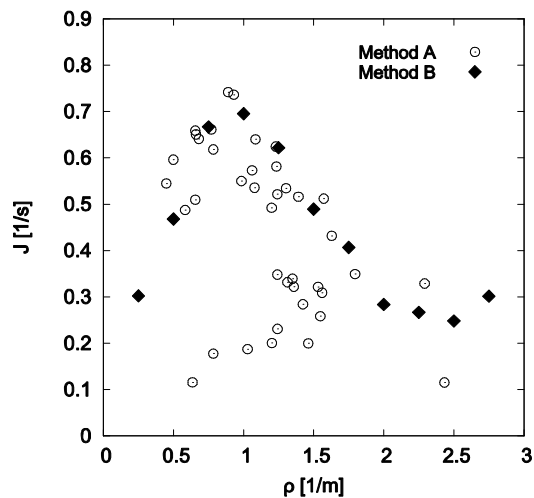

**Fig. 2. Fundamental diagram of single-file movement determined by different measurement methods.**

*Method A*: **Direct measurement of the flow and velocity at a cross-section. The density is calculated via** $\rho = <J>_{\Delta t} / <v>_{\Delta t}$. **Method B: Measurement of the density and velocity at a certain time point averaged over space. The flow is given by** $J = \rho <v>_{\Delta x}$.

The deviations of the results obtained by the two methods depend on the fact that the velocity distributions measured at a certain location and averaged over time do not necessarily conform with velocity distributions measured at a certain point of time averaged over space. This is an important point for a quantitative macroscopic validation procedure comparing experimental data with simulation results.

We have recently performed a set of experiments with up to 250 persons under well controlled laboratory conditions. Great emphasis was given to the method of data recording by video technique and careful preparation of the experimental setups. A more general discussion of the experimental setups, the definition of the objectives and some preliminary results are presented in (Seyfried et al., 2009).

## 2.3. Flow at bottlenecks

In applications, one of the most important questions is how the capacity of a bottleneck increases with increasing width. Studies of this dependence can be traced back to the beginning of the last century (Dieckmann, 1911; Fischer, 1933) and are still discussed controversially. Intuitively, a stepwise increase of capacity with the width appears to be natural, especially in the case of lane formation. If these lanes are independent, i.e. pedestrians in one lane are not influenced by those in others, the capacity can only increase when an additional lane can be formed.

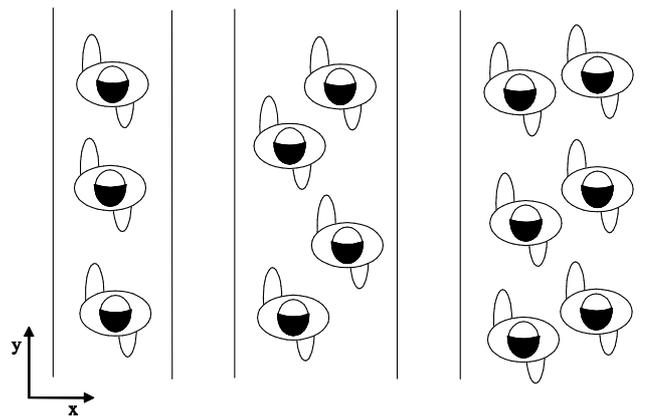

**Fig. 3. Zipper effect with continuously increasing lane distances: The distance in the walking direction decreases with increasing lateral distance. Density and velocities are the same in all cases, but the flow increases continuously with the width of the section.**

In contrast, the study (Seyfried et al., 2009a) found that the lane distance increases continuously as illustrated in Fig. 3. Moreover this continuous increase leads to a very weak dependence of the density and velocity inside the bottleneck on its width.

To find a conclusive judgment whether the capacity grows continuously with the width the results of different laboratory experiments (Seyfried et al., 2009a; Müller, 1981; Muir et al., 1996; Nagai et al., 2006; Kretz et al., 2006) are compared in (Seyfried et al., 2009a), see Fig. 4. The data by (Muir et al., 1996) from airplane evacuations seem to support the stepwise increase of the flow with the width. They show constant flow values for $b > 0.6$ m. But the independence of the flow over the large range from $b = 0.6$ m to $b = 1.8$ m indicates that in this special setup the flow is not restricted by the bottleneck width. Thus all collected data for flow measurements in Fig. 4 are compatible with a continuous and almost linear increase with the bottleneck width for $b > 0.6$ m.

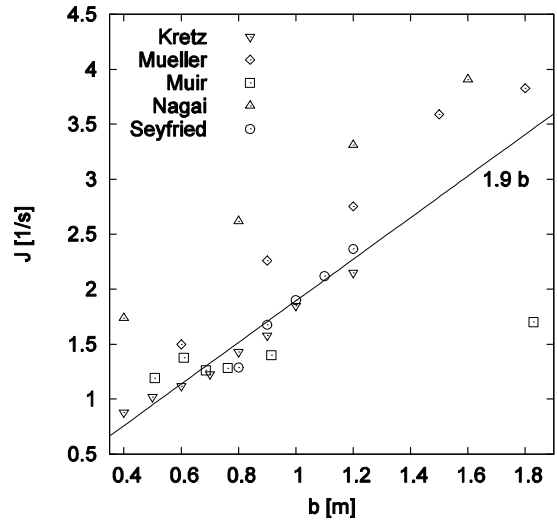

Fig. 4. Influence of the bottleneck width on the flow. Experimental data (Seyfried et al., 2009a; Müller, 1981; Muir et al., 1996; Nagai et al., 2006; Kretz et al., 2006) for different bottleneck types and initial conditions. All data are taken under laboratory conditions where the test persons are advised to move normally.

Surprisingly the data in Fig. 4 differ considerably in the values of the bottleneck capacity. In particular the flow values of (Nagai et al., 2006) and (Müller, 1981) are much higher than the maxima of empirical fundamental diagrams. It appears that the exact geometry of the bottleneck is of only minor influence on the flow while a high initial density in front of the bottleneck can increase the resulting flow values. This leads to the interesting question how the bottleneck flow is connected to the fundamental diagram. General results from nonequilibrium physics show that boundary conditions only select between the states of the undisturbed system instead of creating completely different ones (Popkov and Schütz, 1999). Therefore it is surprising that the measured maximal flow at bottlenecks can exceed the maximum of the empirical fundamental diagram. These questions are related to the common jamming criterion. Generally it is assumed that a jam occurs if the incoming flow exceeds the capacity of the bottleneck. In this case one expects the flow through the bottleneck to continue with the capacity (or lower values). The data presented in (Winkens et al.,2009) show a more complicated picture. While the density in front of the bottleneck amounts to $\rho \approx 5.0(\pm 1)$ m$^{-2}$, the density inside the bottleneck tunes around $\rho \approx 1.8$ m$^{-2}$.

# 3. Models for pedestrian dynamics

## 3.1. Model classes

A large variety of models for pedestrian dynamics has been proposed, ranging from macroscopic approaches based on analogies with hydrodynamics to rather sophisticated multi-agent models (Bandini, et al., 2004; Kukla et al., 2003) taking into account e.g. details of the decision-making processes of the individuals (for a review, see e.g. (Schadschneider et al., 2009).
There are several ways of classifying the different modelling approaches:
- microscopic vs. macroscopic description,
- discrete vs. continuous variables (space, time, state),
- deterministic vs. stochastic dynamics,
- rule-based vs. force-based interactions,
- high vs. low fidelity description.

Molecular dynamics based models are microscopic approaches where the agents are represented as self-driven objects moving in a continuous space. One example is the Social Force Model (Helbing and Molnar, 1995; Helbing et al., 2000). Interactions are given by (generically deterministic) repulsive forces with remote action, but this does not adequately take into account all relevant features. Modifications are necessary, e.g. to account for the empirically observed velocity-density relation (Seyfried et al., 2006; Seyfried et al., 2005), especially the increasing step size at high walking speeds and other observations (Lakoba et al. 2005).

Cellular automata, e.g. (Fukui and Ishibashi 1999; Muramatsu et al., 1999; Klüpfel et al., 2000; Blue and Adler, 2000; Burstedde et al.,2001) are discrete in space, time and state variable. Usually the space discretisation is determined by the space requirement of a person in a dense crowd ($\approx 40 \times 40$ cm$^2$). A timestep is then identified with the reaction time of a pedestrian and is this of the order of a few tenths of a second. CA models have become quite popular recently, probably because they allow for an intuitive definition of the dynamics in terms of simple rules. These are usually stochastic and specified by transition probabilities $p_{ij}$ to one of the neighbouring cells $(i, j)$ (Fig. 5). The transition probabilities for a specific particle are determined by the position of other particles in its vicinity. More realistic models like the floor field model also take into account further influences, e.g. the infrastructure.

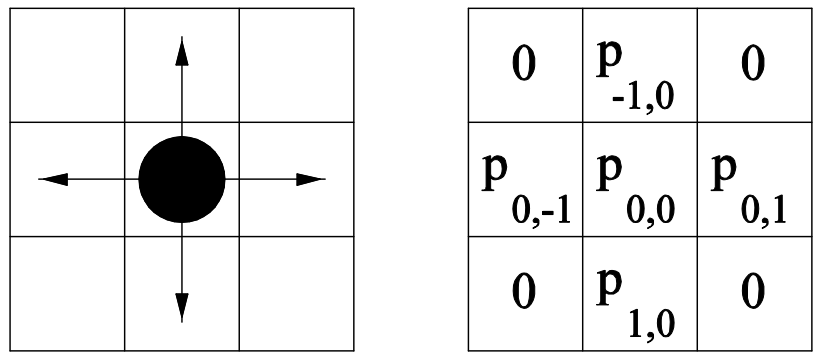

**Fig. 5.** Definition of the transition probabilities $p_{ij}$ for a von Neumann neighbourhood.

## 3.2. Floor field model

The floor field model (Burstedde et al.,2001; Kirchner et al.,2002; Kirchner et al.,2003a) is perhaps the most flexible CA approach as it incorporates the three relevant factors that determine the motion of a pedestrian in a unified way. These factors are
- the desired direction of motion, e.g. to find the shortest connection;
- interactions with other pedestrians;
- interactions with the infrastructure (walls, doors, etc.).

This is achieved by taking inspiration from the motion of ants which is based on process of *chemotaxis* (Hölldoblery et al.,1990; Chowdhury et al.,2005), a chemical form of communication. Introducing a kind of *virtual chemotaxis* allows to translate effects of longer-ranged interactions into purely local ones. Ants deposit so-called pheromones to mark their paths. A similar mechanism is used in the floor field model to take into account the mutual interactions of pedestrians and those with the infrastructure. The virtual pheromones generate *floor fields* which enhance transition probability in the direction of stronger fields.

However, the main factor for the determination of the transition probabilities is the preferred walking direction and speed. This information is encoded in the so-called *matrix of preference*

$M_{ij}$. Its matrix elements are directly related to observable quantities, namely the average velocity and its fluctuations (Burstedde et al.,2001).

These basic probabilities are modified by two discrete *floor fields*, $D$ and $S$. The field strengths $D_{ij}$ and $S_{ij}$ at site $(i, j)$ modify the transition probabilities in such a way that a movement in the direction of higher fields is preferred. The *dynamic floor field $D$* represents a virtual trace left by moving pedestrians. Similar to the process of chemotaxis, this trace has its own dynamics, namely diffusion and decay which lead to the broadening and dilution of the trace with time. The *static floor field $S$*, also called potential in other models, does not change in time and reflects the infrastructure. In the case of the evacuation processes, the static floor field describes the shortest distance to an exit door. The field value increases in the direction of the exit such that it is largest for door cells. An explicit construction of $S$ can be found in (Kirchner et al.,2002; Nishinari et al.,2004).

The full transition probability to cell a neighbouring cell $(i, j)$ is then given by

$$p_{ij} = NM_{ij} e^{k_S S_{ij}} e^{k_D D_{ij}} (1-n_{ij}). \qquad (1)$$

The occupation number $n_{ij}$ is 0 for an empty and 1 for an occupied cell where the occupation number of the cell currently occupied by the considered particle is taken to be 0. The factor $N$ ensures the normalization $\sum_{(i,j)} p_{ij} = 1$ of the probabilities. $k_S$ and $k_D \in [0, \infty[$ are sensitivity parameters that control the relative influence of the fields $D$ and $S$. They have a simple interpretation. The coupling $k_D$ to the dynamic floor field controls the tendency to follow in the footsteps of others, which is often called *herding*. In the absence of a matrix of preference, $k_S$ determines the effective velocity of a single agent in the direction of its destination.

The floor field model is one of the most sophisticated approaches for the description of pedestrian dynamics. Several simpler CA models have been proposed (Schadschneider et al., 2009) which do not include floor fields. There transition probabilities $p_{ij}$ are constant and depend only on the current local configuration in the neighbourhood of a particle. However, these models are not able to reproduce the details of the empirically observed behaviour.

## 3.3. Fundamental diagram of the floor field model

The fundamental diagram incorporates information about the relevance of mutual interactions of the agents at finite densities. Here, due to hindrance effects, their velocity will be reduced compared to the free walking speed.

Typically fundamental diagrams are obtained empirically and theoretically for quasi-one-dimensional motion, e.g. along a corridor. Lateral motion is possible, but will mainly occur to avoid collisions. Since the motion in this situation consists basically of weakly-coupled one-dimensional lanes, where only a few lane changes occur, it is not surprising that the fundamental diagrams are very similar to that of the strictly one-dimensional variant of the model. The latter exhibits the symmetry $J(\rho) = J(\rho_{max})$ where $\rho_{max}$ is the density where the flow vanishes (often normalized to $\rho_{max} = 1$). Thus the function $J(\rho)$ is almost symmetric around the density $\rho_{max} / 2$ with deviations coming from lane changes induced by collision avoidance or fluctuations. A typical fundamental diagram obtained for the basic version of the floor field model (corresponding to $v_{max} = 1$) is shown in Fig. 6.

The comparison with the empirical results of Sec. 2 shows that the observed asymmetry of the fundamental diagram is not reproduced correctly. The origin of this discrepancy is the restriction to models with nearest-neighbour interactions which do not capture essential features like the dynamic space requirement of the agents which depends on their velocity

(and thus density).

Modifications of the floor field model (Kirchner et al.,2004; Kretz and Schreckenberg, 2007) take this effect into account. Here motion is not restricted to nearest-neighbour cells. This is equivalent to a motion at different instantaneous velocities $v = 0,1,...,v_{max}$ where $v$ is the number of cells an agent moves. Then $v_{max} = 1$ corresponds to the case where motion is allowed only to nearest neighbours. Note that different extensions of this type are possible, depending on how one treats crossing trajectories of different agents (Kirchner et al.,2004). But in all cases, the fundamental diagrams become more realistic since the maximum of the flow is shifted towards smaller densities with increasing $v_{max}$ (Fig. 6), in accordance with the empirical observations.

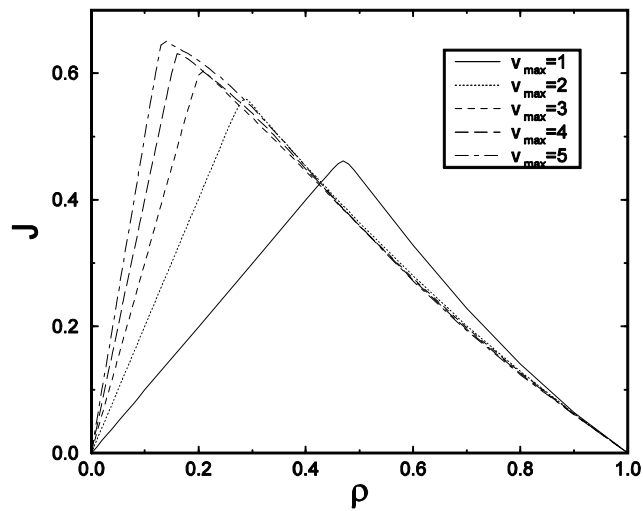

**Fig. 6. Fundamental diagrams of the floor field model for $v_{max} = 1,...,5$. The maximum of the flow is shifted towards smaller densities for increasing $v_{max}$.**

Another modification that appears to be necessary to reproduce empirical observations concerns the size of the cells. The cell size generically chosen corresponds to the space requirement of a single agent, i.e. $40 \times 40$ cm. Since an agent occupies exactly one cell this does not allow to model overlapping lanes like those occurring in the zipper effect (see Sec. 2). This indicates that the cell size used in simulations should be smaller, so that e.g. an agent occupies $2 \times 2$ cells (Kirchner et al., 2004).

## 3.4. Conflicts and friction

Usually, cellular automata and multi-agent models are based on discrete time dynamics which is realized in computer simulations through a synchronous (parallel) updating scheme. This is important for many applications since it implies the existence of a well-defined timescale that can be used for calibration and thus allows e.g. for quantitative predictions. This update scheme leads to inherent problems if at the same time an exclusion principle has to be satisfied, i.e. if a site can not be occupied by more than one particle at the same time. Such restriction is natural for any particle-hopping model related to transport or traffic problems, e.g. intracellular transport, highway traffic and pedestrian dynamics (Chowdhury et al., 2005; Chowdhury et al., 2000).

In this case *conflicts* occur where two or more particles try to move to the same destination cell within the same timestep (Fig. 7). Since multiple occupations are not allowed, a procedure to resolve these conflicts has to be defined (Burstedde et al., 2001). Conflicts might appear to be undesirable effects which reduce the efficiency of execution of simulations and should therefore be avoided by choosing a different update scheme. However, it turns out that they are important for a correct description of crowd dynamics (Kirchner et al., 2003a), especially in clogging situations near bottlenecks. In real life this often leads to dangerous situations and injuries during evacuations. Although conflicts are local phenomena they can have a strong influence on global quantities like evacuation times. In the following we will show how the inclusion of conflicts improves the realism of the model dynamics.

In real life, conflict situations often lead to a moment of hesitation where the involved agents hesitate before trying to resolve the conflict. This reduces on average the effective velocities of all involved particles. This is taken into account in a modification of the floor field model by introducing a probability $\mu$ at which movement of *all* particles involved in the conflict is denied, i.e. all pedestrians remain at their site (see Fig. 7). This means that with probability $1-\mu$ one of the individuals moves to the desired cell. This effect is called *friction* and $\mu$ *friction parameter* since it has similar consequences as contact friction, e.g. in granular materials. It does not reduce the velocity of a freely moving particle and effects only show up in local interactions.

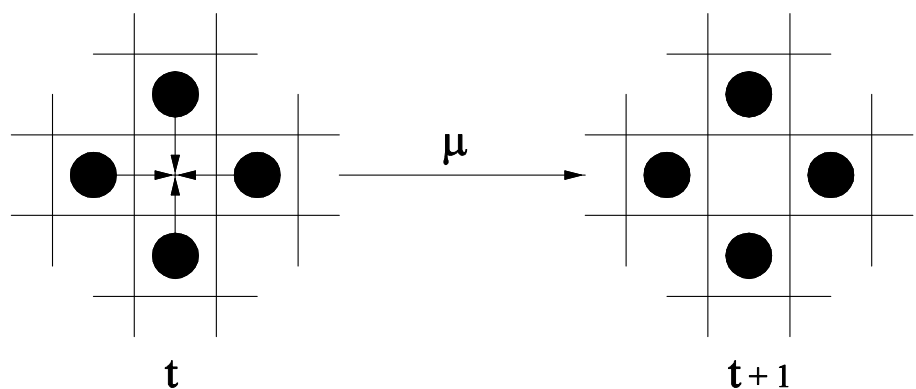

**Fig. 7. In a conflict several particles try to move to the same destination cell at the same time. The friction parameter $\mu$ determines the probability that such a conflict is not resolved and no particle will move.**

Friction has a substantial influence on the dynamics in large density situations. For example it leads to a faster-is-slower effect (Helbing et al., 2002; Helbing et al., 2000) where an increase of the free velocity of the pedestrians does not lead to reduced evacuation times in the presence of friction (Kirchner et al., 2003a). This can be understood since for larger velocities even for relatively low densities jams will form at the exit. In such a situation many conflicts occur and thus large friction has a strong influence on the evacuation time (Fig. 8). Another characteristic effect that is caused by friction is the bursty behaviour of the outflow.

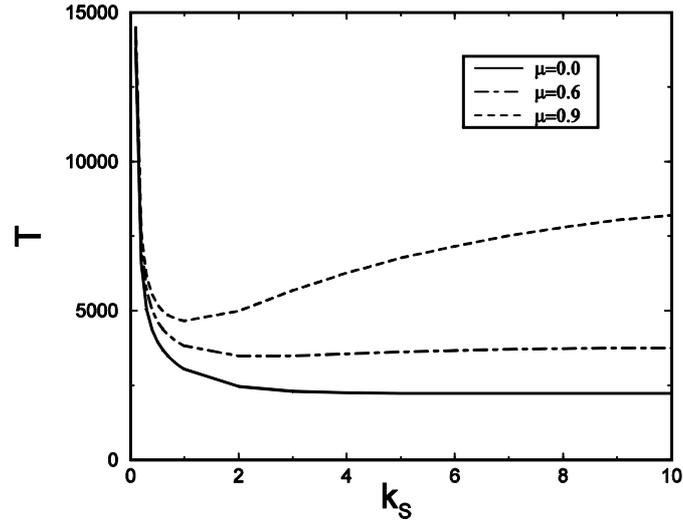

**Fig. 8.** Evacuation time as function of the walking speed (controlled by the parameter $k_S$) for different friction strengths $\mu$. For $\mu = 0.9$ a faster-is-slower effect is observed, i.e. the minimal evacuation time is not found for the largest walking speed (corresponding to $k_S \to \infty$).

Another empirical result which shows the relevance of friction effects for the modelling of pedestrian dynamics comes from the study of evacuation times from airplanes as function of the exit width and the motivation level of passengers (Muir et al., 1996). It is found for narrow exits non-competitive (cooperative) passenger behaviour leads to faster egress whereas for wider exits competitive behaviour is advantageous (Fig. 9).

These findings can be reproduced by the floor field model if friction effects are included (Kirchner et al.,2003b) [43]. Competitive behaviour is then described by a large walking speed (controlled by the parameter $k_S$) and large friction effects due to strong hindrance in conflict situations. Cooperation on the other hand corresponds to small speed and friction.

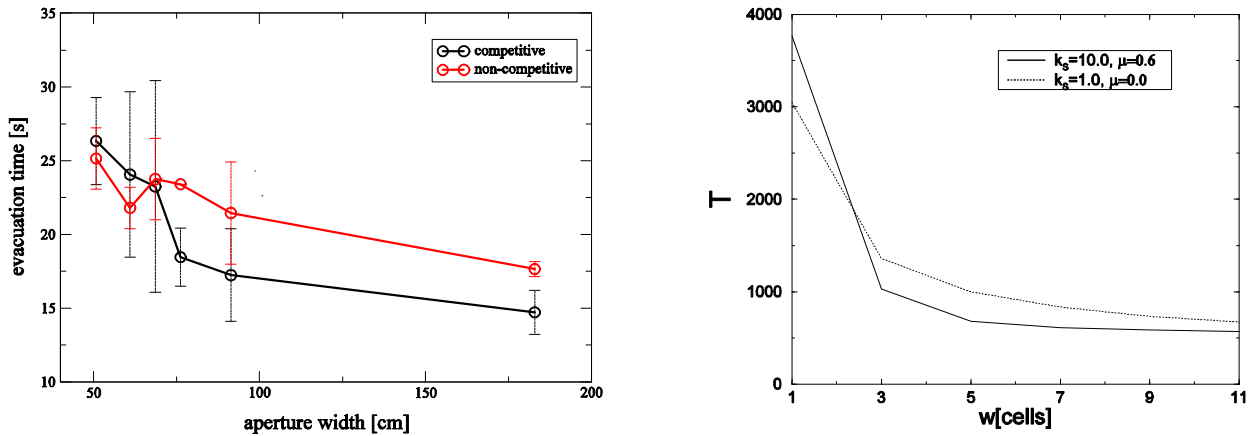

**Fig. 9.** Left: Empirical egress time as function of the door width for competitive and non-competitive behaviour (from [18]); Right: Simulation results based on the floor field model including friction effects.

## 4. Conclusions

We have discussed several aspects of the validation of models for pedestrian and crowd dynamics. A major problem is the lack of reliable and reproducible empirical data where even for the most essential quantities like the capacity there is currently no consensus. This is very unsatisfactory and a serious obstacle in the validation and calibration of the models which is of extreme importance for most applications, especially in the area of safety analysis.

Furthermore we have discussed various modelling approaches, focussing on a special cellular automaton model, the floor field model. It is not only relatively simple and intuitive, but also flexible enough to allow for calibration once the empirical situation has improved. One example is the fundamental diagram which indicates that an extension beyond nearest-neighbour interactions is necessary. We have also discussed the relevance of conflicts and frictions effects as indicated also by experiments. These effects can easily be incorporated in CA approaches like the floor field model which shows the flexibility of this model class.

## Acknowledgments

We thank our collaborators, especially the members of PedNet (`www.ped-net.org`) for helpful discussions.